\documentstyle[preprint,aps]{revtex}
\input epsf
\epsfverbosetrue

\begin{document}
\draft
\preprint{hep-ph/yymmdd}

\title{Coherence Effects in Neutrino Oscillations}

\vskip .5in

\author{Ken Kiers$^a$, Shmuel Nussinov$^b$ and Nathan Weiss$^{a,b,c}$}

\vskip .2in

\address{$^a$Department of Physics, University of British 
Columbia,\\ Vancouver,
British Columbia, Canada V6T 1Z1 \\~\\$^b$Department of Physics,
Tel Aviv University \\ Ramat Aviv, Tel Aviv, Israel \\~\\$^c$Department
of Particle Physics, Weizmann Institute of Science \\ Rehovot, Israel 76100 }

\maketitle

\vskip .3in

\begin{abstract}
We study the effect of coherent and
incoherent broadening  on neutrino oscillations
both in vacuum and in the presence of matter (the MSW effect).
We show under very general assumptions that it is not possible
to distinguish experimentally neutrinos produced in some region of space as
wave packets from those produced in the same region of space
as plane waves with the same energy
distribution.
\end{abstract}

%\pacs{ }
\newpage

\section{Introduction} {\label{sec:1}}
Neutrino oscillations have been the subject of intense theoretical and 
experimental research. To date there is no evidence for oscillations in 
terrestrial neutrino beams. The deficit of solar neutrinos can be explained
by neutrino oscillations with a (mass)$^2$ difference 
$\Delta m^2 \sim 10^{-6} {\rm eV}^2$ together with the enhancement
of these oscillations as the neutrinos pass through the sun by the MSW
effect\cite{ref:solar}.
There are also  hints of possible neutrino
oscillations with $\Delta m^2 \sim 10^{-2} {\rm eV}^2$
in atmospheric neutrino experiments\cite{ref:atmos}.

Approximately twenty-five years ago an interesting suggestion was
made for probing even lower values of $\Delta m^2$ using the
$~3\%$ annual variation of the earth--sun distance\cite{ref:suggest}.
In this case, rather than simply observing a net average decrease
in the electron neutrino intensity by an amount ${\rm sin}^2(2\theta)/2$
one could observe the actual oscillations of the 
electron neutrino flux. The idea is to use the $\nu_e$'s
from  $e^-$ capture on Be
\equation
	^7Be+e^-\rightarrow ^7Li+\nu_e 
	\label{eq:1}
\endequation
which results in a neutrino energy $E_\nu\sim .86~{\rm MeV}$ with a
small energy spread.
Thus if the neutrino oscillation length
\equation
	L_{\rm osc} = \frac{4\pi E_\nu}{\Delta m^2} 
	\label{eq:2}
\endequation
is within one or two orders of magnitude of the variation 
$\Delta R \sim 5\times 10^{11}~{\rm cm}$
of the earth--sun distance then, depending on the value of 
${\rm sin}^2(2\theta)$, it may be possible to  see the neutrino oscillations
provided $\Delta m^2$ is in the range 
$10^{-9}-10^{-11} {\rm eV^2}$.

One of the most essential ingredients in making the above scenario
work  is that the spread in energy $\Delta E$
of the neutrino ``beam'' is
not too wide. This is especially true in this case since
$R/\Delta R \gg 1$.
If $\Delta E$ is too large then by the time the neutrinos arrive at the earth 
the oscillation patterns for
neutrinos of different energies get sufficiently out of phase
to wipe out any
potentially observable oscillations. This results simply in 
a decrease of the total $\nu_e$ intensity by an amount 
${\rm sin}^2(2\theta)/2$. A coherence length
$L_{\rm max}$ is usually defined as the distance at which a
neutrino of energy $E$ has undergone one oscillation
more than a neutrino of energy $E+\Delta E$. This coherence length
is given by
\equation
	L_{\rm max}=\frac{4\pi E^2}{\left(\Delta m^2\right)\Delta E}
	= L_{\rm osc}\left(\frac{E}{\Delta E}\right) 
	\label{eq:3}
\endequation
and the total number of complete oscillations will be
\equation
	N_{\rm max}=\frac {L_{\rm max}}{L_{\rm osc}} 
	\label{eq:4a}
\endequation
Thus when $\Delta E/E$ is larger than about $1/30$ we can no longer
observe the oscillations and a narrow energy
range $\Delta E$ is therefore required.

The argument above assumed that the energy spread of the neutrino
beam is incoherent in origin in the sense that it is due
to slightly different energies of various neutrinos.
The main origin of this energy spread $\Delta E$ is that the continuum
electrons which are captured by the $Be$ have an energy
spread $\Delta E_e \sim kT$ which translates into a similar
spread $\Delta E_\nu \sim kT$ of the emerging neutrino energies.
Another slightly smaller contribution to $\Delta E_\nu \equiv \Delta E$
originates from the different Doppler shifts due to thermal nuclear
velocities (relative to the line of sight) -- the analog of the well
known Doppler broadening in atomic spectroscopy\cite{ref:1}.

``Coherent broadening'' -- namely the
quantum mechanical spread $\delta E$ of a {\bf single} neutrino
can also lead to the loss of the oscillation pattern\cite{ref:2}.
The well known natural line width in atomic spectroscopy:
\equation
	\delta E \sim \Gamma \sim \left( \tau_{\rm decay}\right)^{-1} 
	\label{eq:5}
\endequation
is an example of coherent broadening. The finite lifetime $\tau_{\rm decay}$
of the level interrupts the classical emission of the wave--train
and limits the size of the wave packet $\delta x$ to:
\equation
	\delta x = c\tau_{\rm decay}
	\label{eq:4}
\endequation
with the momentum $(\delta P=\delta E/c)$ and the configuration
space $(\delta x)$ widths being inversely related to each other
via a Fourier transform of a Lorentzian to an exponential.

Another example of coherent broadening is the collisional broadening
(also known as the ``pressure broadening'') of the neutrino line.
It stems from the interruption of coherent emission by collisions
of the emitting atoms. The corresponding wave packet size is given
by an analog of Eq.  (\ref{eq:4}) but with $\tau_{\rm decay}$
replaced by $t_{\rm collision}$ -- the effective time interval between
``relevant collisions''. This (nuclear) collisional broadening 
effect has been extensively studied as the major contributor
to the loss of coherence in neutrino
oscillation. There have been various estimates of the strength
of the effect leading to estimates of the size of
$\delta x =ct_{\rm collision}$\cite{ref:2},\cite{ref:3},\cite{ref:1},
\cite{ref:4}.

A third contribution to the coherent broadening which 
we believe is likely to 
contribute even more to the energy spread $\delta E$ of the neutrino
wave packet is the small size of the wave packets of the captured 
electrons. Since the K electron ionization energy in Berillium
$E_{\rm ion}=Z^2R_{\rm y}=16R_{\rm y}\sim 220~{\rm eV}$ is small in comparison
with the thermal $kT \sim {\rm keV}$ energy, the capture in reaction 
(\ref{eq:1}) is primarily that of continuum electrons. An electron wave packet
of size $\delta_e$ will traverse the (point like) nucleus in a time
\equation
	\delta t = \frac {\delta_e}{v_e}
	\label{eq:7}
\endequation
where $v_e$ is the velocity of the electron. Because the weak interaction 
underlying the capture process (\ref{eq:1}) is local, the time available for 
the $\nu_e$ emission is $\delta t$ and the size of the outgoing $\nu_e$ 
wave packet emitted with velocity $c$ ($c=1$ in our units) will be 
\equation
	\delta_\nu=\frac{\delta_e}{v_e} 
	\label{eq:8}
\endequation
The thermal kinetic energy of a typical electron is 
$\frac{1}{2} m_e v_e^2\sim \frac{3}{2}kT$. Thus
\equation
	v_e=\sqrt{\frac{3kT}{m_e}}\sim .08 
	\label{eq:9}
\endequation
It remains only to estimate the appropriate wave packet size $\delta_e$
to be used in Eq. (\ref{eq:7}). The electrons suffer many random
collisions in the hot core which tend
to localize the wave function and reduce the wave packet size.
If the only information available is that the electrons are in thermal
equilibrium then $\delta_e$ is expected to be of the order of the thermal 
wave length:
\equation
	\delta_e\sim \frac{2\pi}{m_e v_e} \sim \frac{2\pi}
	{\sqrt{3m_e kT}} 
	\label{eq:10}
\endequation
which then leads to a neutrino wave packet size
\equation
	\delta_\nu=\frac{2\pi}{\delta E_\nu} \sim 6\times 10^{-8} {\rm cm} 
	\label{eq:11}
\endequation
This $\delta_\nu$ is {\bf smaller} (and the corresponding incoherent
broadening is {\bf larger})
than all previous estimates.

The three mechanisms described above all lead to the conclusion that
neutrinos are emitted in the sun as  wave packets with 
a rather small size $\delta_\nu$ corresponding  to
``coherent broadening'' of the neutrino line by an amount
$\delta E \sim 2\pi/\delta_\nu$.
This coherent broadening also leads to the loss of the oscillation
pattern\cite{ref:2} after a 
coherence length $L_{\rm coh}$ which is precisely
equal to the coherence length $L_{\rm max}$ derived in 
Eq. (\ref{eq:3}). This result can be derived technically by
decomposing the wave packet into plane waves of energy $E$ with a probability
distribution
\equation
	P(E)=\vert\Psi\left(E\right)\vert^2 
	\label{eq:12}
\endequation
and repeating the discussion leading to Eqs. (\ref{eq:3}) and (\ref{eq:4a}).
This leads to identical conclusions but with $\Delta E$ replaced by
the energy spread $\delta E$  given by:
\equation
	\left(\delta E\right)^2=\int dE~P(E)\left(E^2-\bar E^2\right)
\endequation  

There is, however, a simple intuitive explanation
for how the oscillations are lost in terms of the wave packet
of the neutrino in configuration space\cite{ref:2}. 
Consider an electron neutrino wave packet which is emitted at $t=0$ 
from the solar core. At $t=0$ the $\nu_e$ can be written as a superposition
of two wave packets with identical shape corresponding to the mass
eigenstates $\vert \nu_1\rangle$ and $\vert \nu_2 \rangle$.
\equation
	\vert\nu_e(t=0)\rangle={\rm cos}(\theta)\vert\nu_1\rangle
	+{\rm sin}(\theta)\vert\nu_2\rangle 
	\label{eq:13}
\endequation
This initial wave packet will quickly spread in the directions
$(x,y)$ perpendicular
to the direction of motion but the spreading in the direction
of motion ($z$) is negligible due to Lorentz contraction effects.
Due to the different mass of the $\nu_1$ and  $\nu_2$ their wave 
packets travel with a different (group) velocity
\equation
	\Delta v=v_2-v_1=\frac{\Delta m^2}{2E^2} 
	\label{eq:14}
\endequation
Thus after a time $t$ has elapsed and the neutrino has traveled
a distance $r\sim t$ from the source the two wave packets move with
respect to each other by an amount
\equation
	\Delta r = \Delta v t \sim \frac{\Delta m^2}{2E^2} r 
	\label{eq:15}
\endequation
Neutrino oscillations are simply the ``beating'' of the two wave 
packets as they slide relative to each other by $\Delta r=\lambda$ with
\equation
	\lambda=\frac{2\pi}{E} 
	\label{eq:16}
\endequation
the wavelength of the neutrino. The oscillation length of Eq. (\ref{eq:2})
is then recovered as:
\equation
	L_{\rm osc}=\left\{{\rm value~of~r~for~which}~\Delta r=\lambda\right\}
	=\frac{\lambda}{\Delta v} =\frac{4\pi E}{\Delta m^2} 
	\label{eq:17}
\endequation
The total number of possible neutrino oscillations is simply the
total number of wavelengths within the wave packet,
$N_{\rm max}=\delta_\nu/\lambda=E/\delta E$. 
After this number of oscillations
the two wave packets do not overlap at all and all oscillations are lost.
Thus the coherence distance $L_{\rm coh}$ which is the maximum distance
over which we see oscillations is given by
\equation
	L_{\rm coh}=N_{\rm max}L_{\rm osc} =\left(\frac{E}{\delta E}\right)
	L_{\rm osc} 
	\label{eq:18}
\endequation
which is precisely the result of Eqs. (\ref{eq:3}) and (\ref{eq:4a}) for
the case of incoherent energy broadening.
Indeed once
$\Delta r$ is greater than the size $\delta_\nu$ of the wave packet
the $\nu_1$ and the $\nu_2$  will have completely separated 
spatially.  We would thus expect that they will not interfere
when interacting locally with an electron or nucleus in a detector.

The main aim of this paper is to study whether the two
effects discussed above namely
the incoherent versus the coherent broadening can be distinguished.
They are clearly distinct physical phenomena which can be controlled
(at least in principle) at the {\bf source}. In an Atomic
Physics analog the Doppler broadening
can be controlled relative to the natural line width by
adjusting the temperature of the system or by confining the atoms to a narrow 
channel transverse to the line of sight\cite{ref:5}.
The more interesting question is:  Can we distinguish
these effects at the detector? In this paper we shall show that
in all physically interesting situations the answer is ``no''.
We shall discuss some simple cases in which this answer is 
clear and then we shall prove some general theorems which will show
that under a wide variety of physically attainable situations these
two effects cannot be distinguished.

\section{Coherent versus Incoherent Broadening}

Our goal in this section is to see whether one can distinguish
an incoherent ensemble of plane waves with a mean energy $E$ and
an energy spread $\Delta E$ from an ensemble of wave packets 
each with the same mean energy $E$ and the
same energy width $\delta E=\Delta E$. Before proceeding we should make 
one point clear. Even in the ``incoherent'' case in which we have
an ensemble of plane waves these waves certainly do not have an
infinite extent in the $z$ direction (the direction of motion).
In fact even if we took each ``plane wave'' (with an energy in the
{\rm MeV} range) to have an energy uncertainty of the order of
$10^{-5}{\rm eV}$
(which is certainly a great underestimate for the 
solar neutrino case) the corresponding wave packet
would still be only of the order of a {\rm cm} in size!! Thus when discussing
``plane waves'' we are in fact referring to wave packets which are much
larger than those discussed in the case of coherent broadening but much
smaller than any macroscopic scales in the problem.

\subsection{An Example \label{ss:eg}}

Our aim will be to show that the two broadening effects discussed
above cannot be distinguished.
We begin with a concrete suggestion for distinguishing these effects
and we then show
what goes wrong with this suggestion.

Let us suppose that we were able to measure the energy of a neutrino
with a precision $\epsilon$ which is
much better than $\delta E=\Delta E$. We then 
expect that for an incoherent beam of neutrinos with energies
in a range $\Delta E$ about $E$ we could recover the oscillations
by measuring the neutrino energy to the precision $\epsilon \ll \Delta E$.
By plotting the observed neutrino count as a function of
\equation
	r^\prime=r\frac{\bar E}{E}
\endequation
we should see
oscillations  up to a new distance
\equation
	L_\epsilon=(E/\epsilon)L_{\rm osc}> L_{\rm max}
	\label{eq:18a}
\endequation
with no loss of statistics.
Note that $\Delta E$ is replaced by
$\epsilon$ in Eq. (\ref{eq:3}).

If, on the other hand, we began with
a {\rm wave packet} with energy spread $\delta E$ then, at a distance
larger than $L_{\rm coh}$ (Eq. (\ref{eq:18})), the wave packet of 
the $\nu_1$ and the $\nu_2$ are completely separated  and one might
naively expect that there will be no oscillations even if the energy
could be measured more accurately.

This argument turns out to be wrong and we can understand what
goes wrong in a very intuitive way. If we choose to measure the energy
very accurately (to an accuracy $\epsilon$) we require a time
$t\sim 1/\epsilon$ to make this measurement. If, during this time $t$,
the second wave packet arrives at the detector then we will once more
see the oscillations.  The condition for recovering the oscillations
is therefore 
\equation
	t \sim 1/\epsilon  > \Delta r \sim \frac{\Delta m^2}{2E^2} r
	\label{eq:19} 
\endequation
where $\Delta r$ is the distance between the wave packets and $r$ is 
the distance from the source (Eq. (\ref{eq:15})). The oscillations thus
persist up to a new distance $L_\epsilon^\prime$ which is the
value of $r$ for which Eq. (\ref{eq:19}) breaks down.
\equation
	L_\epsilon^\prime=(E/\epsilon)L_{\rm osc}
	\label{eq:20}
\endequation
which is precisely the same as the result (\ref{eq:18a}) obtained for
the incoherent neutrino beam.

This behavior of the coherent beam is analogous to what occurs
for a high Q oscillator hit by two successive pulses. 
The first pulse (in our case the $\nu_1$ beam)
comes along and sets the oscillator in motion. It then continues to
oscillate for a time $t\sim 1/\epsilon$ during which time the
second pulse (in our case the $\nu_2$ beam) arrives and
causes the oscillator to be further excited. In this way coherence is
maintained between the $\nu_1$ and the $\nu_2$ beams even when
they are spatially separated. What happens is that the accurate measurement
of the energy picks out the plane wave in the wave packet which has
existed coherently through both pulses.

Our main goal will be to understand how general the above result is.
In other words, to what extent is it true that an ensemble of plane
waves will give the same result as wave packets. Although there were some
initial attempts to distinguish these processes it is now widely 
believed that they are indistinguishable. Our goal in this paper
is to prove some theorems which clarify the conditions under which
this is true and to show how general the result is.

\subsection{Measuring Observables which Commute with Momentum}

Before discussing the most general situation we review here the 
proof that the coherently and incoherently broadened neutrino
beams lead to the same total rate and energy distributions
for both $\nu_e$'s and $\nu_\mu$'s.

\subsubsection{Oscillations in Vacuum \label{sss:vac}}
Let us consider two cases representing two possible {\bf electron} 
neutrino beams\footnote{They could, of course, be any linear combination
of electron and muon neutrinos. Electron neutrinos were chosen
for definiteness only.} leaving some region of
the sun at time $t=0$. In \underline{case a} we have an incoherent
mixture of neutrinos each of which is a nearly ideal plane wave
(with some extremely small energy spread $\delta E_{\rm pw} <<< \delta E$). 
In this mixture the probability of finding a neutrino
of energy $E$ is given by some probability distribution $P(E)$
which is centered about some energy $E_0$ with a width $\Delta E$.
In \underline {case b} all the neutrinos come with the same quantum state.
This state is a wave packet with amplitude  $\Psi(E)$ for a plane wave
component of energy $E$. We choose this amplitude so that the probability
distribution $\vert\Psi(E)\vert^2$ precisely matches the distribution
$P(E)$ of \underline{case a}. Consequently the widths of the two distributions
are also equal: $\delta E=\Delta E$. In this section,
for simplicity, we shall treat the plane
waves of \underline{case a} as ideal plane waves with
$\delta E_{\rm pw}=0$.

At $t=0$ the wave function for the \underline{case b} is given by:
\equation
	\vert \psi(t=0)\rangle= \sum_{p,i}\alpha_{p,i}\vert p,i\rangle
	\label{eq:21}
\endequation
where the sum (which is actually an integral)
is over momenta $p$ in the $z$ direction (the direction of
motion) and over mass eigenstates $i=1,2$ and the $\alpha_{p,i}$ are 
chosen to give an electron neutrino with the appropriate wave function at
$t=0$. Since the $\vert p,i\rangle$ are eigenstates of the
Hamiltonian, 
at a later time $t$, the wave function is given by
\equation
	\vert \psi(t)\rangle= \sum_{p,i}\alpha_{p,i}
	{\rm e}^{-i\epsilon_p^{(i)}t}\vert p,i\rangle
	\label{eq:22}
\endequation
where $\epsilon_p^{(i)}=\sqrt{p^2+m_i^2}$ is the energy of $\nu_i$
with momentum $p$. 

Suppose now that at time $t$ we measure an observable $Q$ which 
{\it commutes with the momentum operator}. $Q$ may, for example, be
the total number of electron neutrinos in some range of momenta.
This is, in fact, the most common kind of measurement which can be made.
In this case $Q$ has only diagonal matrix elements in momentum space.
Therefore the expectation value of $Q$ at time $t$ is given by:
\equation
	\langle \psi(t)\vert Q \vert \psi(t) \rangle =
	\sum_p \left[ 
	\left(\sum_j \alpha^*_{p,j}{\rm e}^{i\epsilon_p^{(j)}t}
	\langle p,j \vert \right) Q \left( \sum_i \alpha_{p,i}
	{\rm e}^{-i\epsilon_p^{(i)}t} \vert p,i \rangle \right) \right]
	\label{eq:23}
\endequation
The expression inside the square brackets is precisely the  expression
for the expectation value $\langle Q\rangle_p$ of $Q$  for a plane wave
which has a total weight (i.e. normalization)
$\vert\alpha_{p,1}\vert^2+\vert\alpha_{p,2}\vert^2$ and a {\it relative}
amplitude 
$\alpha_{p,1}$ and $\alpha_{p,2}$ for $\nu_1$ and $\nu_2$ respectively
at $t=0$.
Thus 
\equation
	\langle Q\rangle=\sum_p \langle Q\rangle_p
	\label{eq:24}
\endequation
which is precisely the result one obtains for 
the incoherent beam of \underline{case a}. Thus the measurement of
any observable which commutes with momentum yields the same result
for \underline{case a} and \underline{case b}.

Although this result may seem trivial it is in fact rather powerful.
>From this result we can verify the result claimed in Sec. \ref{sec:1}
that if we use the variation in the earth--sun 
distance to look for oscillations
in the neutrinos from $^7Be$ both the coherent and the incoherent neutrino
beams give the same oscillation pattern. A--priori the above theorem
is not applicable since the experiment involves measuring the spatial
dependence of the neutrino flux which involves the use of an operator
which does not commute with momentum. This is however an
example for which the conversion of spatial to temporal 
dependence can be done reliably. Thus, although we measure the spatial
variation in the neutrino flux, we can compute the temporal dependence
of this flux by computing, for example, the total number of 
electron neutrinos with a given energy as a function of time. 
This estimate will be
reliable since, as discussed previously,
even the ``plane wave'' packet
is still extremely small (certainly much less than 
a {\rm cm} in size) relative to the relevant astronomical scales.

\subsubsection{Oscillations in Matter \label{sss:matter}}

The above proof that an ensemble of plane waves cannot be distinguished,
at the detector, from an ensemble of wave packets with the same energy
distribution can be extended to the case of neutrino oscillations in 
matter (the MSW effect)\cite{ref:msw}. To this end imagine that at 
$t=0$ an electron neutrino is produced (at the origin) in matter in which 
the density of electrons (along the direction of motion of the neutrino)
is given by $\rho_e(z)$. (This is of course an approximation in which
we neglect variations of the density in the transverse directions.)
The ``vacuum eigenstates'' $\vert \nu_1\rangle$
and $\vert\nu_2\rangle$ are no longer eigenstates of this system.
Instead one can find new eigenstates of the Hamiltonian which include
the full spatial variation of the density. These eigenstates
will of course no longer be momentum eigenstates.  For relativistic neutrinos
one should, in principle, solve the Dirac Equation but for the present
discussion since spin is not a crucial variable
it suffices to consider the 
Klein--Gordon equation\cite{ref:kg}:
\equation
	\left(-\left(\frac{\partial}{\partial t} +i 
	{\rm A_m}\right)^2+ \nabla^2\right)
	\left(\matrix{\psi_e\cr\psi_\mu}\right) =
	{\rm M_0^2} \left(\matrix{\psi_e\cr\psi_\mu}\right)
	\label{eq:25}
\endequation
where ${\rm M_0}$ is the vacuum mass matrix and the matrix $A_m$ 
accounts for the 
effect of charged--current scattering of the $\nu_e$
off the electrons in the medium:
\equation
	M_0^2=\left(\matrix{m_1^2\cos^2\theta+m_2^2\sin^2\theta&
	\left(m_1^2-m_2^2\right)\sin\theta\cos\theta\cr
	\left(m_1^2-m_2^2\right)\sin\theta\cos\theta&
	m_1^2\sin^2\theta+m_2^2\cos^2\theta}\right) , ~~~
	A_m=\left(\matrix{\sqrt{2}G_F\rho_e(z)&0\cr 0&0}\right)
	\label{eq:26}
\endequation

The eigenstates of this
system with energy $E$ will no longer be eigenstates of $p_z$
($p_x$ and $p_y$ are assumed to be zero) but
will be labeled by some other parameter which we call $\gamma$.
We shall call these eigenstates 
\equation
	\vert \gamma , 1 \rangle ~~~{\rm and} ~~~ \vert\gamma,2\rangle
	\label{eq:27}
\endequation
In the regions of space where the density vanishes
these eigenstates will behave as plane waves\footnote{It may in fact be 
a superposition of an incoming and an outgoing plane wave if there
is reflection.} with some momentum
$p_{\rm vac}(\gamma)$.  They will correspond to  vacuum mass eigenstates of
the system. In a region of space in which the electron density is
nonzero but nearly constant the eigenstates $\vert\gamma,i\rangle$
will again be nearly plane waves but now corresponding to the
usual neutrino eigenstates {\bf in matter}.

Suppose now that at $t=0$ we prepare an electron neutrino in a state
described by some rather narrow wave 
packet $\vert \psi(t=0)\rangle$. (This is 
\underline{case b} of Sec. \ref{sss:vac}.)
At $t=0$, in analogy with Eq. (\ref{eq:21}), 
this state can be expanded in the eigenstates
$\vert \gamma,i\rangle$ described above 
\equation
	\vert \psi(t=0)\rangle=\sum_{\gamma,i}\alpha_{\gamma,i}
	\vert\gamma,i\rangle
	\label{eq:28}
\endequation
We now allow the state to propagate to a later time $t$ .
At this later time the state is given by
\equation
	\vert \psi(t)\rangle=\sum_{\gamma,i}\alpha_{\gamma,i}
	{\rm e}^{-i\epsilon_\gamma^{(i)}t}\vert\gamma,i\rangle
	\label{eq:30}
\endequation
where $\epsilon_\gamma^{(i)}$ is the energy of the state
$\vert \gamma , i\rangle$.  In any reasonable case the 
size of the wave packet
at time $t$ will  be much smaller than the scale of 
variations in the
electron density. (This is especially true if the measurement is made
in vacuum.)  Thus Eq. (\ref{eq:30}) amounts to an expansion in
the momentum eigenstates of the neutrinos in matter with density
$\rho$ equal to the density at the location of the wave packet.
Every $\gamma$ corresponds to some 
momentum $p(\gamma)$ which depends on the
density $\rho$. Thus
\equation
	\vert \psi(t)\rangle\simeq \sum_{p,i}\hat\alpha_{p,i}
	\vert p,i; \rho\rangle
	\label{eq:29}
\endequation
where for any given value of $p$ and the corresponding value of $\gamma$
the coefficients $\hat\alpha_{p,1}$ and  $\hat\alpha_{p,2}$ are 
linear combinations  of $\alpha_{\gamma,1}$ and $\alpha_{\gamma,2}$.

Now suppose that at time $t$ we measure some operator $Q$ which commutes
with (the $z$ component of the) momentum. The off--diagonal matrix 
elements of $Q$ vanish in the momentum basis. Thus $Q$
will have only diagonal matrix elements between the various
$\vert p,i;\rho\rangle$ in Eq. (\ref{eq:29}). Since each of these
$\vert p,i;\rho\rangle$ corresponds to one of the energy eigenstates
$\vert\gamma,i\rangle$ it follows that
the expectation value of $Q$ is given by 
\equation
	\langle \psi(t)\vert Q \vert \psi(t) \rangle \sim
	\sum_\gamma \left[\left( \sum_j \alpha^*_{\gamma,j}
	{\rm e}^{i\epsilon_\gamma^{(j)}t}
	\langle \gamma,j \vert\right) Q\left( \sum_i \alpha_{\gamma,i}
	{\rm e}^{-i\epsilon_\gamma^{(i)}t} 
	\vert \gamma,i \rangle\right)\right]
	\label{eq:31}
\endequation
This expression is analagous to Eq. (\ref{eq:23}) in Sec. \ref{sss:vac}.
Each term in the sum is precisely the result which we would have obtained
for the expectation value $\langle Q\rangle_\gamma$ of $Q$  
for a state which was initially in an approximate momentum eigenstate
corresponding to $\gamma$ but with total weight
$\vert\alpha_{\gamma,1}\vert^2+\vert\alpha_{\gamma,2}\vert^2$ 
and a {\it relative} amplitude 
$\alpha_{\gamma,1}$ and $\alpha_{\gamma,2}$ 
for $\nu_1$ and $\nu_2$ respectively.
(Recall that the realistic
plane waves are actually extremely narrow on the scale of the
density variations.)
Thus 
\equation
	\langle Q\rangle=\sum_\gamma \langle Q\rangle_\gamma
	\label{eq:31a}
\endequation
which is precisely the result one obtains for 
the incoherent beam of \underline{case a}. 
There is thus no difference between the wave packet 
(\underline{case b}) and the plane wave (\underline{case a}) ensemble 
{\it even in matter} when only operators which commute with momentum
are measured.

\subsection{Unrealistic Measurements which CAN Identify Wave Packets}

>From the above proof it seems that a keen measurement which combines
a measurement of both position and momentum information might
be able to distinguish an ensemble of wave packets from 
an ensemble of plane waves.
The simplest way of doing this would, however, require precise knowledge
of the point of origin and the time of origin of the wave packet.
Suppose, for example, that we {\bf knew} that all the wave packets
in our ensemble 
(\underline{case b}) were centered at the origin ($z=0$) precisely 
at time $t=0$. Suppose also that in the alternative scenario
(\underline{case a}) we also knew that each (nearly ideal) plane
wave (which is still a wave packet but with a much larger spatial
extent than that of \underline{case b})  in the ensemble was
centered at the origin at $t=0$. 
Under these assumptions about our previous
knowledge and
by a careful timing measurement at the earth to  determine
the duration of the neutrino pulse we {\bf could} distinguish
the two cases. (In fact in \underline{case b} there may be
two separated pulses.)  This scenario is, of course,
totally unrealistic and we shall see below that if we allow for an
uncertainty in the location of the initial 
packets it again becomes impossible to distinguish the two cases
by {\it any} measurement at the earth.

There is another scenario under which it is clearly possible to
distinguish the two cases. Suppose we have a detailed theory
for the production mechanisms of the two cases which lead to
some different observable at the source. Suppose, for example,
that the position or momentum distributions for the two cases
are expected to differ. Then clearly such information can be
used to decide which mechanism is producing the neutrinos (or,
more realistically, which mechanism dominates). However in the case of
level broadening we have no such information. Both the energy and the
position distributions are expected to be roughly the same. The question
which we are asking is: Assume we are given two ``sources'' of neutrinos 
(or production mechanisms) with the {\bf same} position ($z$)
and momentum ($p$) distributions. Is it possible to tell 
by measurements at the detector which of the two ``sources''
produced these neutrinos?

\subsection{General Theorem }\label{sss:eng}

This question can be set up more precisely as follows:
Consider the following two modifications of the scenarios
\underline{case a} and \underline{case b} discussed above:

In \underline{case A} we have a nearly ideal plane wave which
is actually a wave packet of a fairly large size $\Delta z$.
(Recall that $\Delta z$ will typically be much less than a {\rm cm}!).
We imagine an ensemble of such ``plane waves'' each of which 
has a nearly precise momentum (in the $z$ direction)
centered about $p_0$ with a spread $\delta p$.
Assume that each plane wave has exactly the same spatial location.
(This is precisely the \underline{case a} above.)

In \underline{case B} we have an ensemble of wave packets. 
Each wave packet has a spatial size $\delta z$ which is much
smaller than $\Delta z$ and a corresponding momentum spread
$\delta p = 1/\delta z$ which is precisely equal to the 
$\delta p$ of \underline{case A}. Up to this point this looks exactly
like \underline{case b} above except we now allow each wave
packet in our ensemble to be, at $t=0$, at a different spatial 
location. We assume that the wave packets are produced in precisely the
same region $\Delta z$ in which the neutrinos of \underline{case A}
are produced with precisely the same $z$ distribution.\footnote{In a
realistic situation both the ``plane waves'' of {\underline{case a}}
and the wave packets of \underline{case b} will be distributed over 
a region of space much larger than $\Delta z$.  In both cases
this excess spread is incoherent. It is thus sufficient to prove
our result for the case when the wave packet is distributed in $z$
by the size $\Delta z$ of the plane wave of \underline{case a}.}
The two cases are shown pictorially in Figure I.

All the above information is given to the experimenter together with
the additional information that the $z$ and $p$ distributions for
both cases are equal at $t=0$. The question is:
With only this information can the experimenter distinguish 
with any experiment the cases A and B above?

Intuitively one might guess that the answer is ``yes''. There should
be some way to tell if we are dealing with wave
packets or with (almost)
plane waves! But in fact the answer is ``no''!! No experiment
can distinguish the above two cases. 

The most general proof of this statement would proceed as follows:

{\bf Step 0.} Choose  values for $\delta z=1/\delta p \ll \Delta z$ 
and for the mean momentum $p_0$ which were defined above.

{\bf Step 1.} Begin with an arbitrary (smooth) 
but fixed expression for the wave 
function of the nearly ideal plane wave of \underline{case A}.
The only constraints on this wave function will be that it is centered
(say) at the origin, that its spread in position is (a fairly large)
$\Delta z$  with a correspondingly tiny
spread in momentum about some momentum $p$.
Then consider an ensemble of such states each with a different momentum
$p$. Choose an arbitrary but fixed distribution for these momenta.
The constraint on this distribution is that it is centered about the
momentum $p_0$ with the given width $\delta p$.

{\bf Step 2.} 
Construct the Density Matrix for the ensemble described in Step 1 above.

{\bf Step 3.} 
One must now prove that it is always possible to construct the following,
seemingly completely
different ensemble, which, nonetheless, yields a density matrix 
identical to the one obtained in Step 2 above.
We  first construct a wave packet
which is centered at some location $z$. We are free to choose
the form of the wave function  with the only constraint that
its spread in position is approximately equal to
$\delta z \ll \Delta z$ with a corresponding
momentum spread $\delta p=1/\delta z$. We then  construct
an ensemble of such wave packets and choose a  distribution of
locations $z$ with the only constraint that this distribution 
be centered at the origin
with a spread in position approximately equal to $\Delta z$. 

The claim is that we can {\bf always} choose the distributions
in Step 3 so that the density matrix for Step 3 is identical to that
of Step 2. This then implies that any measurement at all
which is done on the two ensembles {\bf at any time $t$} gives the
same result!! We also claim the converse of this theorem namely that
given a ``wave packet'' ensemble constructed as in Step 3 it
is always possible to find a ``plane wave'' ensemble as constructed
in Step 1 with the same density matrix.

Note how the mass eigenstates $\nu_1$ and $\nu_2$ appear nowhere
in the above discussion. The reason for this and, in our opinion,
the power of this proof is that it relies entirely on properties
of the system at $t=0$ at which time the state is a pure
$\nu_e$ state.

\subsubsection{Illustration in the Simplest Case}

We can show the essence of the proof by the following simple
example. We model the wave packet (of \underline{case B}) by a 
superposition of only two 
momentum eigenstates $\vert p_1\rangle$  and $\vert p_2\rangle$.
In Step 1 above we imagine having the state $\vert p_1\rangle$
with probability $\vert\alpha\vert^2$ and the state $\vert p_2\rangle$
with probability $\vert\beta\vert^2$; 
($\vert\alpha\vert^2+\vert\beta\vert^2=1$). 
The density matrix for this system is simply
\equation
	\vert\alpha\vert^2 \vert p_1\rangle\langle p_1 \vert
	+ \vert\beta\vert^2 \vert p_2\rangle\langle p_2 \vert
	\label{eq:32}
\endequation
For implementing Step 3 we may construct an analogue of 
wave packets at two different
locations as an ensemble consisting of these two states
with equal probability:
\equation
	\vert \psi_\pm\rangle =\
	\alpha \vert p_1\rangle \pm \beta\vert p_2\rangle
	\label{eq:33}
\endequation
The density matrix in this case
\equation
	\vert \psi_+ \rangle\langle \psi_+\vert~+~\vert \psi_- \rangle\langle 
	\psi_-\vert
	\label{eq:34}
\endequation
is precisely the same as the density matrix for \underline{case A}
in Eq. (\ref{eq:32}). 

This completes the proof in this simple case.

\subsubsection{Gaussian Distributions}

One case in which the Steps 0-3 above can be carried out explicitly
is when all distributions are Gaussian. 
Thus in Step 1 we choose the ``plane wave'' of momentum
$p$ to have a wave function
\equation
	\vert p; {\rm plane}\rangle = 
	\frac{1}{\left(\sqrt{2\pi}\sigma\right)^{\frac{1}{2}}}\int dl~
	{\exp}\left(-\frac{(l-p)^2}{4\sigma^2}\right)\vert l \rangle
	\label{eq:35}
\endequation
where $\sigma \sim 1/\Delta z$.
We then consider an ensemble of these states with a Gaussian distribution
of momenta $p$
\equation
	\frac{1}{\left(\sqrt{2\pi}\sigma_p\right)}~
	{\exp}\left(-\frac{(p-p_0)^2}{2\sigma_p^2}\right)
	\label{eq:36}
\endequation
where $\sigma_p \sim \delta p = 1/\delta z \gg \sigma$.
The density matrix for this case (\underline{case A})
is given by
\equation
	\rho_A=\frac{1}{\left(\sqrt{2\pi}\sigma_p\right)}
	\frac{1}{\left(\sqrt{2\pi}\sigma\right)}
	\int dl~dl^\prime~ dp ~{\exp}\left(-\frac{(p-p_0)^2}{2\sigma_p^2}
	-\frac{(l-p)^2}{4\sigma^2}-\frac{(l^\prime-p)^2}{4\sigma^2}\right)
	\vert l \rangle\langle l^\prime \vert
	\label{eq:37}
\endequation

We now proceed with Step 3 corresponding to \underline{case B}.
Consider a wave packet with mean momentum $p_0$ centered at 
some location $z$:
\equation
	\vert z; {\rm packet}\rangle = 
	\frac{1}{\left(\sqrt{2\pi}\hat\sigma_p\right)^{\frac{1}{2}}}\int dl~
	{\rm e}^{-ilz}
	{\exp}\left(-\frac{(l-p_0)^2}{4\hat\sigma_p^2}\right)\vert l \rangle
	\label{eq:38}
\endequation
We shall soon see that the correct choice for $\hat\sigma_p$ is  
\equation
	\hat\sigma_p^2=\sigma_p^2+\sigma^2
	\label{eq:39}
\endequation 
which is approximately equal to $\sigma_p=1/\delta z$ as required.
We then consider an ensemble of these states with a Gaussian distribution
of positions $z$ centered at the origin of the form
\equation
	\frac{1}{\left(\sqrt{2\pi}\sigma_z\right)}\int dz~
	{\exp}\left(-\frac{2z^2}{\sigma_z^2}\right)
	\label{eq:40}
\endequation
The correct choice for spread in position, $\sigma_z$, will turn out to be:
\equation
	\sigma_z^2=\frac{\sigma_p^2}
	{\sigma^2\left(\sigma^2+\sigma_p^2\right)}
	\label{eq:41}
\endequation
This $\sigma_z$ is approximately equal to $1/\sigma=\Delta z$ as required.
The density matrix for this situation (\underline{case B})
is given by
\equation
	\rho_B=\frac{1}{\left(\sqrt{2\pi}\hat\sigma_p\right)}
	\frac{2}{\left(\sqrt{2\pi}\sigma_x\right)}
	\int dl~dl^\prime~ dz~
	\left[{\exp}\left(-\frac{2z^2}{\sigma_z^2}
	-\frac{(l-p_0)^2}{4\hat\sigma_p^2}-
	\frac{(l^\prime-p_0)^2}{4\hat\sigma_p^2}\right)\right]
	 ~ {\rm e}^{-i\left(l-l^\prime\right)z}
	\vert l \rangle\langle l^\prime \vert
	\label{eq:42}
\endequation

With the choices we have made for $\hat\sigma_p$ and $\sigma_z$
in Eqs. (\ref{eq:39}) and (\ref{eq:41}) it turns out that the density matrices
$\rho_A$ and $\rho_B$ are {\bf precisely equal}. The calculation is
straightforward and most easily done by 
computing the matrix elements $\langle l\vert\rho_{A,B}\vert l^\prime\rangle$.
In order to compute the matrix elements of 
$\rho_A$ only the integral over $p$ must be done. This is a Gaussian
integral. For $\rho_B$ only  the integral over $z$ must be done.
This is simply the Fourier Transform of a Gaussian. The result is the same
for $\rho_A$ and $\rho_B$ and is given by:
\equation
	\langle l\vert\rho_{A,B}\vert l^\prime\rangle
	=\frac{1}{\left(\sqrt{2\pi}\hat\sigma_p\right)}~
	\exp\left(-\frac{\left(l-p_0\right)^2}{4\hat\sigma_p^2}
	-\frac{\left(l^\prime-p_0\right)^2}{4\hat\sigma_p^2}
	-\frac{\left(l-l^\prime\right)^2\sigma_p^2}{8\sigma^2\hat\sigma_p^2}
	\right)
	\label{eq:43}
\endequation
We thus establish, for the Gaussian case, that the two ensembles
are identical.

\subsubsection{General Proof}

In the Gaussian case described above we did not use the fact
that $\delta z \ll \Delta z$. In the case of a more general shape
for the ``plane wave'' and the wave packet we shall present a proof
which does rely on this approximation. We conjecture that it is
possible to slightly modify the theorem 
\footnote{The modification we have in mind is to relax the unnecessary
restriction that the shape of the wave packet is independent of $z$.
It is reasonable to consider an ensemble of wave packets all of which
have the same width but with slightly different shapes. The same
could be done for the ``nearly plane waves''.} 
so that it will be valid
for general values of $\delta z$ and $\Delta z$ but we do not
have a proof at this time.

We begin again with Step 1 for which we choose a ``plane wave'' of momentum
$p$ to have a wave function
\equation
	\vert p; {\rm plane}\rangle = 
	\int dl~
	f_\sigma(l-p) \vert l \rangle
	\label{eq:44a}
\endequation
where the function $f_\sigma(l-p)$ has a width $\sigma \sim 1/\Delta z$.
We then consider an ensemble of these states with a distribution
of momenta $p$ given by some function $g_{\sigma_p}(p-p_0)$ with a width
$\sigma_p \sim \delta p = 1/\delta z \gg \sigma$.
The density matrix for this case (\underline{case A})
is given by
\equation
	\rho_A=
	\int dl~dl^\prime~ 
	dp ~\left[f_\sigma^*(l^\prime-p)
	f_\sigma(l-p)g_{\sigma_p}(p-p_0)\right]
	\vert l \rangle\langle l^\prime \vert
	\label{eq:44b}
\endequation

We now proceed with Step 3 corresponding to \underline{case B}.
Consider a wave packet with mean momentum $p_0$ centered at 
some location $z$:
\equation
	\vert z; {\rm packet}\rangle = 
	\int dl~ {\rm e}^{-ilz}
	\alpha_{\sigma_p}(l-p_0)\vert l \rangle
	\label{eq:44c}
\endequation
which has {\it approximately} a width $\sigma_p$. 
We then consider an ensemble of these states with a  distribution
of positions $z$ centered at the origin given
by some function $h_{\sigma_z}(z)$ with 
a width $\sigma_z$ which is approximately equal to $1/\sigma=\Delta z$.
The density matrix for this situation (\underline{case B})
is given by
$$
	\rho_B=
	\int dl~dl^\prime~ dz~ \left[
	\alpha_{\sigma_p}^*(l^\prime-p_0)
	\alpha_{\sigma_p}(l-p_0) h_{\sigma_z}(z)
	\right]{\rm e}^{-i(l-l^\prime)z}
	\vert l \rangle\langle l^\prime \vert
$$
\equation
	= \int dl~dl^\prime~ \left[
	\alpha_{\sigma_p}^*(l^\prime-p_0)\alpha_{\sigma_p}(l-p_0) 
	\tilde h_{\sigma}(l-l^\prime) \right]
	\vert l \rangle\langle l^\prime \vert
	\label{eq:44d}
\endequation
where $\tilde h_{\sigma}(l-l^\prime)$ is the Fourier transform of 
$h_{\sigma_z}(z)$ which has a width approximately equal to $\sigma$. 

The requirement that the two density matrices are equal is now simply
stated as:
\equation
	\alpha_{\sigma_p}^*(l^\prime-p_0)\alpha_{\sigma_p}(l-p_0) 
	\tilde h_{\sigma}(l-l^\prime) =
	\int dp ~f_\sigma^*(l^\prime-p)f_\sigma(l-p)g_{\sigma_p}(p-p_0)
	\label{eq:45}
\endequation
(It is now clear why the theorem, as stated, cannot be true in general.
Given arbitrary smooth functions $g_{\sigma_p}$ and $f_\sigma$ with the
restrictions described previously it is certainly not possible, in general,
to find functions $\alpha_{\sigma_p}$ and $\tilde h_\sigma$ which
satisfy Equation (\ref{eq:45}) since the integral in (\ref{eq:45}) will
not always factorize in the required form.) The result is however valid
when the width $\sigma_p$ of $g_{\sigma_p}$ is much larger than
the width $\sigma$ of $f_\sigma$. If $\sigma \ll \sigma_p$ and if the
function $g_{\sigma_p}$ is sufficiently smooth
%\footnote{Specifically,
%we require that $\frac{g_{\sigma_p}'(l-p_0)}{g_{\sigma_p}(l-p_0)}
%=O\left(\frac{1}{\sigma_p}\right)$ for $\vert l-p_0\vert <\sim\sigma_p$.}
$$
	f_\sigma^*(l^\prime-p)f_\sigma(l-p)g_{\sigma_p}(p-p_0)
	~=
$$
\equation
	f_\sigma^*(l^\prime-p)f_\sigma(l-p)
	\sqrt{g_{\sigma_p}(l-p_0)}\sqrt{g_{\sigma_p}(l^\prime-p_0)}
	+O\left(\frac{\sigma}{\sigma_p}\right)
\endequation
Thus 
\equation
	\int dp ~f_\sigma^*(l^\prime-p)f_\sigma(l-p)g_{\sigma_p}(p-p_0)
	\sim
	\sqrt{g_{\sigma_p}(l-p_0)}\sqrt{g_{\sigma_p}(l^\prime-p_0)}
	\times \int dq ~ f_\sigma^*(q)f_\sigma(l-l^\prime +q)
\endequation
Thus if we identify the function $\alpha_{\sigma_p}$ with the square root
of $g_{\sigma_p}$ and the function $h_{\sigma_z}(z)$ with the square of the Fourier
transform of $f_\sigma(p)$ then the equality in Eq. (\ref{eq:45}) is
satisfied to order $\delta z/\Delta z$ as required.

\subsection{Consequences}

Although the result proven above is not entirely general it is sufficient
for all cases of practical interest. The reason for this is
that we have actually proven three things. The result that measurements
which commute with momentum could not distinguish coherent from incoherent
broadening was completely general and did not depend on the shape of
the wave packet nor on its width. Secondly the proof that for Gaussian
wave packets the two effects could not be distinguished with {\bf any}
measurement was also general and it did not depend on the width of
the Gaussians. Thirdly our extension of the proof to arbitrary wave
packet shapes was possible in the limit $\delta z \ll \Delta z$. 
A practical attempt to distinguish the two mechanisms of broadening
would likely begin with a theoretical calculation which
assumes  Gaussian wave packets for simplicity.
Furthermore it would likely
compare the wave packets to  actual 
plane waves for which $\Delta z\rightarrow \infty$. 
We have {\bf shown} that any such attempt
is doomed to failure. We conjecture that the result is more general
so that for an arbitrary shape of wave packet it is possible to
find an ensemble of nearly plane waves which mimic its behavior exactly.

An interesting corollary to the result proven in the previous
section is that one cannot tell, on an event by event basis,
whether one has a wave packet or a ``plane wave''. The proof is
as follows: Suppose it were possible, on an event by event basis,
to distinguish a wave packet from a plane wave. It would
then be trivially possible to distinguish the cases A and B above
since in one case we are presented with a plane wave and
in the other case with a wave packet. In fact in just one event
we would know with which case we are dealing. But, as we saw
in the previous section we cannot do this since the density
matrices for the two cases are identical. It follows that
no such determination can be made on an event by event basis.

This result does not contradict the recent work of 
several authors\cite{ref:protect} on the ability 
to measure the wave function
of a single particle via a 
``protective measurement''. There are at least two requirements for
such a measurement to be possible. The first is that the system
needs an energy gap so that successive (soft) measurements
keep the particle in the same state. The second requirement
is that it is known a--priori that the system is in an eigenstate of 
the Hamiltonian. Thus, for example, it is in principle
possible to measure the ground state wave function of a typical
atom even on a single atom but, if we do not know whether the
atom is in an eigenstate of the Hamiltonian or in a superposition
of eigenstates then this cannot be determined on a single atom.
An argument very similar to that in the previous paragraph can
be used to prove this result. In our case 
neither of these conditions are satisfied. We do not have a 
gap and we certainly are not in an eigenstate of the Hamiltonian
when we are dealing with a wave packet and/or we start
with a pure flavor state such as a $\vert\nu_e\rangle$.

The theorem presented in the previous section also provides a general
tool for understanding how the size of a Quantum Mechanical
wave packet affects  physical results in various circumstances.
In fact  the style of our proof 
which relies on the
{\it initial} properties of the system rather than on the 
details of its time evolution is extremely useful. 
There have been several instances in which either
careless approximations or faulty logic have lead
to conclusions which disagree with our very general result.
To illustrate this point imagine, instead of using our general
proof, that we evolve each of the two ensembles
to a later time $t$ and {\bf then} compared them.
We must, by our theorem, get the same density matrix for each ensemble.
But in doing this calculation we might make  
several approximations to simplify the calculation. 
We might, for example, neglect the longitudinal 
spreading of the wave packet. 
It turns out that even when this spreading is negligible
compared to the size of the wave packet it has a significant
effect on the final density matrices and we would find
significant differences between the two ensembles. 
We know from our theorem that this cannot be the case. 
Indeed when the effect of
longitudinal spreading is included all results computed with 
$\rho_A$ and $\rho_B$ agree.

\section {Summary and Conclusions}{\label{s5}}

The main focus of this paper was the question of our ability experimentally
(even in principle) to distinguish incoherent broadening of a 
neutrino line  (such as the $^7Be$ solar neutrino line) from coherent
broadening of such a line. Of particular interest was whether these
two types of broadening would have different effects on neutrino 
oscillations and the MSW effect.
We began by identifying processes
which contribute to these mechanisms of broadening. 
Coherent broadening results from several processes including
the natural width of the emitting nucleus, pressure broadening caused 
by collisions of this nucleus and the finite size of the wave packet
of the captured electron. We argued that this last process leads
to the smallest estimate for the spatial size of the neutrino wave
packet ($\sim 6\times 10^{-8}${\rm cm}). 
Incoherent broadening results mainly
from the thermal energy spread of the captured electron as well as from
the Doppler shift due to the thermal motion of the emitting nucleus.

We then began to present our argument that although the two forms
of broadening were distinct physical processes which could be controlled
at the source they could not be distinguished at the detector. 
We first showed that if the detector had an excellent energy resolution
not only could oscillations due to an incoherent ensemble of (nearly)
monoenergetic neutrinos be restored but 
oscillations of a coherent neutrino beam could also be restored despite
the physical separation of the $\nu_1$ and the $\nu_2$ at the detector.
We then proved that the measurement of any operator which
commuted with momentum could never distinguish a wave packet from 
a plane wave. We extended the proof of this result to the case in which
the neutrino propagates in matter (the MSW effect). 

The next stage was to show that if we had no a--priori knowledge of
any difference in the properties of the coherent versus the incoherent
neutrino ``beams'' there was $no$ measurement which could distinguish
them. Our method was to show that it was possible to construct
two ensembles, one corresponding to ``nearly plane waves'' and the
other to wave packets which had the same density matrix at $t=0$.
This would imply that the density matrices were equal at all later 
times and that no measurements could distinguish the two cases.
 We presented a complete proof in the case of Gaussian wave packets
by showing 
that the density matrix at the source for an ensemble of plane waves
with a given (Gaussian) energy distribution was {\bf equal} to that
of an ensemble of wave packets each with a much narrower $z$ distribution
but distributed, incoherently, over the same range of positions as
the ``incoherent'' ensemble. We extended this proof to the case
of non--Gaussian wave packets in the limit that the spatial size
of the wave packet was much smaller than the spatial size of the 
``nearly plane wave''.  We conjectured that the result is even more general
and that given any ensemble of ``nearly plane waves'' with a given energy
and position distribution we can construct an ensemble of wave packets
which has precisely the same density matrix.

There have been claims in the literature that 
wave packets could give different
results than plane waves with the same momentum distribution. These
differences show up either when the neutrinos are nearly nonrelativistic
or when their momentum distribution is extremely broad so that
$\delta p \sim p$. This of course implies that some of the components
of the neutrino wave function are nonrelativistic and that some of the
neutrinos are moving ``backwards''. In all these cases it is essential
to include the longitudinal spreading of the neutrino wave packet
and to remember that if one calculates the number of neutrinos which
should be observed at some location $z$ one must compute the 
flux of neutrinos which involves the neutrino velocity. If these
cautions are kept in mind one confirms the results of our theorem that
there are no differences between the two scenarios.

Although we have chosen to focus this paper on neutrinos and
neutrino oscillations it is clear that the result is much more
general. It applies to any particle for which the question
of the distinguishability of a wave packet from plane waves is relevant.
Some examples include neutral Kaon oscillations and the effect
of wave packets in scattering theory.

\vskip .2in

\centerline{\bf ACKNOWLEDGEMENTS}
We would like to thank Yakir Aharonov, 
Eric Carlson, Fred Goldhaber and Bill Unruh
for helpful discussions. This work was supported in part by the
Natural Sciences and Engineering Research Council of Canada.
Their support is gratefully acknowledged. Ken Kiers and Nathan Weiss
would like to thank the Weizmann Institute 
and Tel Aviv University for their support
and hospitality.

\begin{figure}[ht]
\epsfysize=5in
\epsfbox[0 40 612 459]{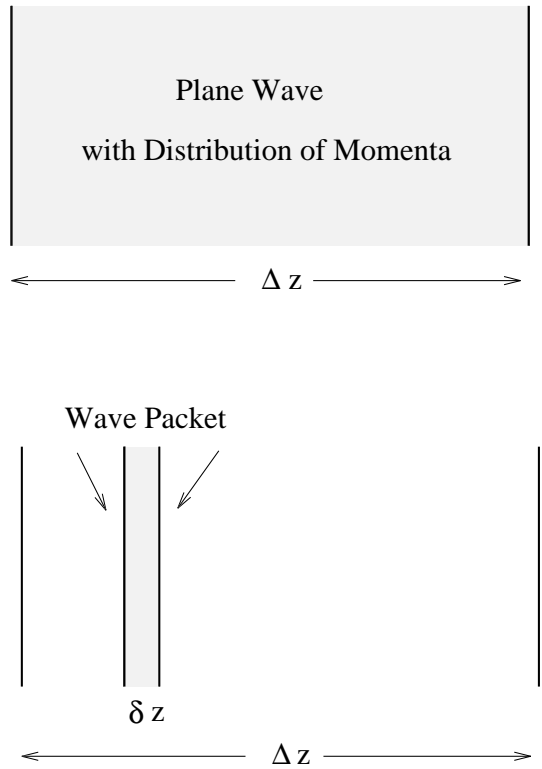}
\caption {Pictorial representation of \underline{Case A} and 
\underline{Case B} described in Sec. IID}
\end{figure}


\begin{references}


\bibitem{ref:solar} Some of the experimental results can be found in:  
	P. Anselmann et al., Phys. Lett. B {\bf 327} (1994) 377;
	R. Davis, in: Proc. Int. Symp. on Neutrino Astrophysics, ``Frontiers
	of Neutrino Astrophysics'', eds. Y. Suzuki and N. Nakamura (Universal
	Acad. Press, Tokyo, 1993) p.47; K. S. Hirata et al., Phys. Rev. D
	{\bf 44} (1991) 2241; V. N. Gavrin, in: `` Proceedings of 
	the XXVI International Conference on High Energy Physics'',  Dallas,
	Texas, 1992, edited by J. Sanford (AIP, New York, 1993).  
	MSW-analyses
	of some of the data can be found in:  
	S. A. Bludman et al., Phys. Rev. D
	{\bf 47} (1993) 2220; P. I. Krastev and S. T. Petcov, Phys. Lett. B
	{\bf 299} (1993) 99; L. Krauss et al., Phys. Lett. B {\bf 299} (1993)
	94.

\bibitem{ref:atmos} Some experimental results can be found in:
	K. S. Hirata et al., Phys. Lett B {\bf 280} (1992) 146;
	D. Casper et al., Phys. Rev. Lett. {\bf 66} (1991) 2561;
	R. Becker-Szendy et al., Phys. Rev. D {\bf 46} (1992) 3720;
	Ch. Berger et al., Phys. Lett. B {\bf 245} (1990) 305;
	M. Aglietta et al., Europhys. Lett. {\bf 8} (1989) 611.
	An analysis of the results is given by E. W. Beier et al.,
	Phys. Lett. B {\bf 283} (1992) 446.  Also see C. W. Kim and
	A. Pevsner, ``Neutrinos in Physics and Astrophysics'',  
	Contemporary Concepts in Physics, Vol. 8, edited by H. Feshbach
	(Harwood Acad. Publishers, Chur, 1993) pp. 151-158.


\bibitem{ref:suggest} B. Pontecorvo, Sov. Phys. -- JETP {\bf 26} (1968) 984;
	V. Gribov and B. Pontecorvo, Phys. Lett. B {\bf 28} (1969) 493.

\bibitem{ref:1} L. Krauss and F. Wilczek, Phys. Rev. Lett. {\bf 55}
	(1985) 122.

\bibitem{ref:2} S. Nussinov, Phys. Lett. B {\bf 63} (1976) 201.

\bibitem{ref:3} A. Loeb, Phys. Rev. {\bf D39} (1989) 1009.

\bibitem{ref:4} C.W. Giunti, C.W. Kim and U.W. Lee, Phys. Lett. B
	{\bf 274} (1992) 87.

\bibitem{ref:5} F. Low, Private Communication

\bibitem{ref:msw} L. Wolfenstein, Phys. Rev. D {\bf 17} (1978) 2369;
	S. P. Mikheyev and A. Yu. Smirnov, Sov. J. Nucl. Phys. {\bf 42}
	(1985) 913; Nuovo Cimento {\bf 9C} (1986) 17.  For an excellent
	review see also T. K. Kuo and J. Pantaleone, Rev. Mod. Phys. 
	{\bf 61} (1989) 937.

\bibitem{ref:kg} This version of the Klein-Gordon equation is similar
	to that encountered in the context of Optical Potential models
	in Nuclear Physics.  See, for example, E. H. Auerbach, D. M.
	Fleming, and M. M. Sternheim, Phys. Rev. {\bf 162} (1967) 1683;
	{\bf 171} (1968) 1781.

\bibitem{ref:protect} Y. Aharonov and L. Vaidman, Phys. Lett. A 
	{\bf 178} (1993) 38; Y. Aharonov, J. Anandan, and L. Vaidman,
	Phys. Rev. A {\bf 47} (1993) 4616.  See also W. G. Unruh, 
	Phys. Rev. A {\bf 50} (1994) 882; W. G. Unruh, Ann. N.Y. Acad. Sci. 
	{\bf 755} (1995) 560.
\end{references}
\end{document}